\documentclass[10.5pt]{article}

\input epsf.sty

\def\lromn#1{\uppercase\expandafter{\romannumeral#1}}

\usepackage{graphicx,amsmath,amssymb}
\def\lromn#1{\uppercase\expandafter{\romannumeral#1}}

\begin{document}

\vspace{2cm}
\begin{center}
\begin{Large}
{\bf Radiative emission of neutrino pair from
nucleus and inner core electrons in heavy atoms}

\end{Large}

\vspace{1cm}
M. Yoshimura and N. Sasao$^{\dagger}$

Center of Quantum Universe, Faculty of
Science, Okayama University \\
Tsushima-naka 3-1-1 Kita-ku Okayama
700-8530 Japan

$^{\dagger}$
Research Core for Extreme Quantum World,
Okayama University \\
Tsushima-naka 3-1-1 Kita-ku Okayama
700-8530 Japan \\

\end{center}

\vspace{5cm}

\begin{center}
\begin{Large}
{\bf ABSTRACT}
\end{Large}
\end{center}

Radiative emission of neutrino pair (RENP) from atomic states
is a new tool to experimentally investigate undetermined
neutrino parameters such as the smallest neutrino mass,
the nature of neutrino masses (Majorana vs Dirac), and
their CP properties.
We study effects of neutrino pair emission either from
 nucleus or from inner core electrons in which
the zero-th component of quark or electron vector current
gives rise to large coupling.
Both the overall rate and the spectral shape of
photon energy are given for a few cases of
interesting target atoms.
Calculated rates exceed those of previously considered target atoms
by many orders of magnitudes.

\vspace{5cm}

Key words

Neutrino mass, Majorana particles, Macro-coherence

\newpage
{\lromn 1}
{\bf Introduction}

\vspace{0.5cm}
Recent developments of neutrino oscillation
experiments have achieved remarkable success:
many elements of the fundamental neutrino mass matrix
have been determined, including
all three mixing angles and
two mass squared differences  \cite{nu oscillation data}.
They however left  undetermined
the absolute scale of neutrino masses
(or equivalently the smallest neutrino mass),
the nature of masses (Dirac or Majorana type),
and their CP properties.
Conventional targets  in ongoing experiments of exploring these
undetermined neutrino properties and parameters
have been nuclei.
Direct measurement of the end point
spectrum of beta decay such as tritium \cite{tritium} and 
(neutrino-less) double beta
decay \cite{nu0 beta} are two main methods to resolve
these outstanding problems.

Some time ago we proposed to use atomic transitions
for improved exploration of undetermined neutrino properties
\cite{my-prd}, \cite{ptep overview}.
The idea is to exploit the fact that atomic level
spacings are much closer to expected
neutrino masses and many
experimental methods are available to manipulate atomic
transitions.
The process we use is atomic de-excitation;
$|e \rangle \rightarrow |g\rangle + \gamma + \nu_i \nu_j$
where $\nu_i, i= 1,2,3$ are neutrino mass eigen states.
By measuring the photon energy spectral shape and determining
locations of six thresholds at
$\omega_{ij} \equiv \epsilon_{eg}/2 - (m_i+m_j)^2/(2\epsilon_{eg}) $
($\epsilon_{eg}$ is the atomic level spacing) associated with
the pair emission $\nu_i \nu_j$,
one can determine all neutrino masses
with precision, if the macro-coherence
we proposed \cite{yst pra} works as expected.
The Majorana vs Dirac distinction is made
possible due to the interference effect
of identical Majorana fermions \cite{my-prd}.

The key idea to enhance otherwise small weak rates
for atomic electrons is the use of coherence, which
may change rates $\propto N$(the number of target atoms)
to rates $\propto N^2$.
A prerequisite for experimental success is thus a
development of macro-coherence, which is triggered by two laser
fields accompanying target polarization \cite{yst pra}.
Macro-coherent radiative emission
of neutrino pair has been called RENP
for brevity.
In our preceding works neutrino
pair emission from valence
electrons has been considered, the emission vertex being M1 
(magnetic dipole) type
(actually the spin current).
The interaction of electron with neutrino
contains both charged W and neutral Z
exchange diagrams,
and the axial vector part
of electron current contributed to this form.

In the present work we examine new
types of neutrino pair emission, emission from core electrons
and nucleus, both arising from zero-th component
of vector current of mono-pole nature.
The relevant mono-pole current counts the number
of constituents, hence one may expect
a large contribution from heavy atoms.
A similar enhancement due to the nuclear mono-pole
current has been used in experiments
that have established atomic parity violation,
\cite{bouchiat 2}, \cite{bouchiat exp}
\cite{commins}, \cite{wieman}.
The nuclear mono-pole 
interaction that gives rise
to largest rates is not sensitive to Majorana
CP phases (but sensitive to Majorana vs
Dirac distinction), while smaller rates of pair emission from valence
electrons has a sensitivity to CP phases.
It seems that for complete determination of
the neutrino mass matrix one needs a variety of
targets, presumably with different technological strategies.

We shall give the photon energy spectrum
of RENP for Cs and Xe.
Alkali atoms are chosen as the simplest
atom to show our fundamental ideas,
and Xe is interesting to leave a room for
a possibility of performing RENP
experiments in gas target.
Different targets have special features
of different merits and demerits.
Further detailed study is necessary
to select the best candidate atoms.

The present work is organized as follows.
In Section {\lromn 2} the important idea
of Coulomb assisted RENP which gives rise to
enhancement by a high power of Z is explained
and formulated.
In Section {\lromn 3} RENP spectral
rate that gives a largest rete is given and some numerical example of the spectral
shape is illustrated.
Finally, we summarize in Section {\lromn 4}.
In two appendices, Section {\lromn 5} and {\lromn 6},
we give rudimentary account of Thomas-Fermi model
used for the estimate of Coulomb integral in heavy atoms
and calculate the phase space integration over neutrino momenta.

\vspace{0.5cm}
{\lromn 2}
{\bf Coulomb assisted neutrino pair emission from
nucleus and core electrons}

\vspace{0.5cm}
The four-Fermi interaction of neutrinos with atomic electrons
and quarks in nucleus is given by
\begin{eqnarray}
&&
H_{w} = \langle n | \int dx^3  ({\cal H}_{2\nu}^e(x)+ 
{\cal H}_{2\nu}^q(x)) | n' \rangle
\,,
\\ &&
 {\cal H}_{2\nu}^e = 
\frac{G_F}{\sqrt{2}} \left( 
\bar{\nu}_e \gamma^{\alpha} ( 1- \gamma_5) \nu_e
\bar{e}  \gamma_{\alpha} ( 1- \gamma_5) e
- \frac{1}{2}
\sum_{i}\bar{\nu}_i \gamma^{\alpha} ( 1- \gamma_5) \nu_i
\bar{e} \gamma_{\alpha} (1- 4 \sin^2 \theta_W - \gamma_5) e
\right)
\,,
\label{2nu vertex}
\\ &&
{\cal H}_{2\nu}^q = \frac{G_F}{\sqrt{2}}
\sum_{i} \bar{\nu_i}\gamma_{\alpha} (1-\gamma_5)\nu_i J_{q}^{\alpha}
\,,
\\ &&
\hspace*{-1cm}
J_{q}^{\alpha} = \frac{1}{2} \left(
\bar{u}\gamma^{\alpha}(1-\gamma_5)u
-\bar{d}\gamma^{\alpha}(1-\gamma_5)d
\right)
- 2 \sin^2\theta_W (\frac{2}{3}\bar{u}\gamma^{\alpha} u 
- \frac{1}{3}\bar{d}\gamma^{\alpha} d)
\,.
\label{quark current}
\end{eqnarray}
As usual, the electron neutrino $\nu_e$ is a mixture of 
three mass eigen-states, $\nu_i$;
$\nu_e = \sum_i U_{ei} \nu_i$.
The neutrino interaction with quarks for RENP
is mediated only by Z-exchange interaction alone.

We shall first consider neutrino interaction with atomic
electrons, arising from the term ${\cal H}_{2\nu}^e$.
Atomic electrons may be treated  non-relativistic
and this gives two main contributions;
the spin vector $e^{\dagger}\vec{\sigma} e$ from
the 4-axial vector current and the mono-pole $e^{\dagger}e$
from the 4-vector current.
For transitions in heavy atoms the mono-pole
contribution from all electrons within the closed shell
is expected to be large, since there are of order Z
electrons unlike a single or a few valence electrons.
Contribution of the vector term $\propto \vec{\sigma}$
cancels among many core electrons.
We shall therefore consider in what follows the mono-pole
weak interaction of the form written in terms of
two component spinor fields,
\begin{eqnarray}
&&
 {\cal H}_{2\nu}^e = \frac{G_F}{\sqrt{2}} e^{\dagger} e \sum_{ij} 
b_{ij}\nu_j^{\dagger}  ( 1- \gamma_5) \nu_i
+ O(\frac{1}{m_e}) 
\,,
\label{2nu vertex 2}
\\ &&
b_{ij} = U_{ei}U_{ej}^* - \frac{1}{2} \delta_{ij} (1- 4 \sin^2 \theta_W)
\,, \hspace*{0.5cm}
1- 4 \sin^2 \theta_W \sim 0.044
 \,.
\end{eqnarray}

We shall  consider the neutrino pair emission from one of 
core electrons in a state $|c \rangle$
and dipole (E1) photon emission from an excited
state $|v'\rangle$, first without Coulomb interaction.
In the non-relativistic perturbation theory 
there are two ways in time sequence in which mono-pole core emission
of vertex $b_{ij}\langle c| c\rangle = b_{ij}$ and 
E1 vertex $\langle v | \vec{d}\cdot\vec{E} | v' \rangle$ are arranged.
When contributions from these two diagrams are added,
they give amplitudes of the  form,
\begin{eqnarray}
&&
b_{ij} \langle v | \vec{d}\cdot\vec{E} | v' \rangle
\left( \frac{1}{\epsilon_{v'} - \epsilon_v - \omega} + \frac{1}{-E_{2\nu}}
\right)
\,,
\label{cancellation without coulomb}
\end{eqnarray}
with $E_{2\nu}$ the total energy of two neutrinos.
Two terms in the bracket are the usual energy denominator
factor in the second order perturbation theory.
The energy conservation for the process
$|v'\rangle \rightarrow |v \rangle + \gamma + \nu_i\nu_j$ gives 
$E_{2\nu} = \epsilon_{v'} - \epsilon_v - \omega$,
hence these two contributions exactly cancel.

Radiative neutrino pair emission from
core electrons thus becomes effective, 
only when it is accompanied by Coulomb
interaction between core electrons and valence electron
which emits a photon.
We shall thus consider the third order perturbation of
Coulomb assisted neutrino pair emission, which has
matrix elements between two anti-symmetrized wave functions
of valence and core electrons (E1 vertex omitted for the moment);
\begin{eqnarray}
&&
\sum_c \langle c | {\cal H}_{2\nu}^e | c \rangle
\langle n, c |\frac{\alpha}{r_{12}}|n',c\rangle
\,, \hspace{0.5cm}
\sum_c \langle n, c | \frac{\alpha}{r_{12}} |n', c \rangle
\langle  c |{\cal H}_{2\nu}^e| c \rangle
\,, 
\end{eqnarray}
where $r_{12}$ is the distance between two electrons.
Quantum number of a single electron
wave function, $c$, refers to one of core electrons, while
$n, n'$ (which may or may not be the same) refers to valence electron.

In performing spatial integration of neutrino emission vertex 
${\cal H}_{2\nu}^e$,
one essentially obtains the integrated electron number density 
of the core, since
the wave vectors of plane neutrino wave functions 
hardly changes within a single atom due to much larger wavelength
of emitted neutrinos.
Hence,
$\langle  c |{\cal H}_{2\nu}^e| c \rangle = $
the weak coupling constants $\times$ two plane wave functions of neutrino pair
at a target site.
The remaining part is Coulomb integral and its exchange integral
between valence and core electrons:
\begin{eqnarray}
&&
\langle n, c | \frac{\alpha}{r_{12}} |n', c \rangle
= \int d^3 r_1 d^3 r_2 \psi^*_n(\vec{r}_1) \psi^*_c(\vec{r}_2)
\frac{\alpha}{|\vec{r}_1 - \vec{r}_2|}\psi_{n'}(\vec{r}_1) \psi_c(\vec{r}_2)
+ ({\rm exchange \; Coulomb \;integral})
\,.
\end{eqnarray}
Exchange Coulomb integral turns out numerically much smaller,
hence is neglected.
We shall use Thomas-Fermi model \cite{atomic physics} for estimate of this
quantity in heavy atoms.
In Appendix we give a basic explanation of Thomas-Fermi model
and how to compute the Coulomb integral in the model.
The result for Coulomb integral is summarized as
\begin{eqnarray}
&&
J_c \equiv 
\sum_c \langle n, c | \frac{\alpha}{r_{12}} |n', c \rangle 
= 
\frac{2^{10/3}}{(3\pi)^{2/3}}
Z^{4/3} \frac{1}{2} \alpha^2 m_e {\cal J}_c
\,,\hspace{0.5cm}
{\cal J}_c \sim 0.23
\,,
\\ &&
J_c^2 \sim 50 Z^{8/3} {\rm eV}^2
\,.
\end{eqnarray}
In the Thomas-Fermi model dependence on the valence
principal quantum numbers, $n,n'$, is weak and we shall ignore it.

We next consider Coulomb assisted neutrino pair emission
from nucleus, which turns out larger than that from core electrons.
(The cancellation without Coulomb interaction works in this case, too,
in much the same way as in eq.(\ref{cancellation without coulomb}). )
The relevant Z-exchange interaction arises from zero-th components
of the quark current (\ref{quark current}), which
is conveniently written in terms of proton and neutron number densities;
\begin{eqnarray}
&&
{\cal H}_{2\nu}^q \sim \frac{G_F}{\sqrt{2}}
\sum_i \nu_i^{\dagger}( 1- \gamma_5) \nu_i
j_{q}^0
\,, 
\\ &&
j_{q}^0 = - \frac{1}{2} j^{0}_n + \frac{1}{2}
(1- 4\sin^2 \theta_W)j^{0}_p 
\,,
\end{eqnarray}
where $j^0_n, j^0_p$ are neutron and proton number densities.
Coulomb assisted pair emission for valence electron
transition, $|n \rangle \rightarrow |n'\rangle$, contains
\begin{eqnarray}
&&
Q_w \sum_i \nu_i^{\dagger} \nu_i \langle n'|\frac{Z\alpha}{r}|n \rangle
\,, \hspace{0.5cm}
Q_w = N - 0.044 Z
\,,
\label{valence-nucleus coulomb}
\end{eqnarray}
where $N,Z$ is the neutron and the proton number of nucleus.
The nucleus is assumed to be a point charge.

Thomas-Fermi model gives an estimate of Coulomb integral of
this type.
Its Z-dependence is given by
\begin{eqnarray}
&&
J_N \equiv
\langle n'|\frac{Z\alpha}{r}|n \rangle \sim
\frac{2^{7/3}}{(3\pi)^{2/3}} Z^{4/3} \alpha^2 m_e {\cal J}_N
\,, \hspace{0.5cm}
{\cal J}_N = 
\int_0^{\infty} dx
\frac{\chi(x)^{3/2}}{x^{1/2}} 
\,.
\end{eqnarray}
Numerically, we find that
\begin{eqnarray}
&&
{\cal J}_N \sim 1.6
\,, \hspace{0.5cm}
(Q_w J_N)^2 \sim 
2.5 \times 10^3
Q_w^2 Z^{8/3} {\rm eV}^2
\,.
\end{eqnarray}
The ratio of two Coulomb integrals,
the one from nucleus to the one from core electrons, is of order,
$50 Q_w^2 $,
thus the pair emission from nucleus dominating the process. 
RENP of some atomic processes however has
no contribution of pair emission from nucleus,
and the pair emission from core electrons may
become dominant.
The enhancement factor of rates from nuclear mono-pole pair emission
is roughly $(Q_w J_N)^2$ divided by squared energy spacing of atomic process.

Thomas-Fermi model overestimates these Coulomb integrals
compared with more precise calculation,
since electrons are distributed more towards the center.
We improved the model following \cite{neuffer and commins} such that the potential is
given by a sum of inner core part of total charge $(Z-1) e$
provided by Thomas-Fermi model and the shielded nuclear
Coulomb potential of $- \alpha/r$.
The non-relativistic Schroedinger equation
was then solved with this potential for a valence electron.
This method gives a value for Ce $J_N$ 
smaller by a factor $\sim 2.5$ than the Thomas-Fermi result.
Nevertheless, we shall use in the rest of this work Thomas-Fermi estimate
for Coulomb integrals for simplicity.

The nuclear mono-pole contribution  is  insensitive to the elements
of neutrino mixing matrix  $U_{ei}$, since its contribution
does not involve W-exchange interaction.

\vspace{0.5cm}
{\lromn 3}
{\bf Spectrum rate of RENP}

\vspace{0.5cm}
The Coulomb assisted neutrino pair emission from
nucleus or core electrons may be combined with
E1 (electric dipole) transition from valence electron.
This is expected to give the largest 
RENP rate.
We shall illustrate calculation of Coulomb assisted
radiative emission of neutrino pair from nucleus,
taking alkali atoms of one valence electron.

\begin{figure*}[htbp]
 \begin{center}
 \epsfxsize=0.5\textwidth
 \centerline{\epsfbox{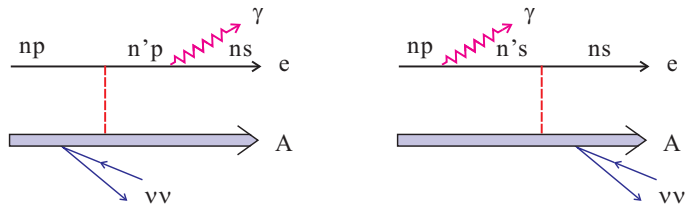}} \hspace*{\fill}
   \caption{RENP diagrams 1 for alkali atoms.
   Red dashed line is for Coulomb interaction between valence electron
and nucleus.
}
   \label{core renp pc 1}
 \end{center} 
\end{figure*}

With the Coulomb assistance, there are six types of diagrams
equally contributing in absolute magnitudes, as shown in
Fig(\ref{core renp pc 1}) $\sim$ Fig(\ref{core renp pc 3}).
There is a partial cancellation of six contributions:
contributions from diagrams of Fig(\ref{core renp pc 1})R
(right diagram) and two of  Fig(\ref{core renp pc 3}) give
a sum of the form,
\begin{eqnarray}
&&
Q_w \langle n' s | \vec{d}\cdot\vec{E} | np \rangle
\langle n's | V_C| ns \rangle
\left(
- \frac{1}{(\epsilon_{n's} - \epsilon_{np} + \omega)
(\epsilon_{np} - \epsilon_{ns} - \omega)} + 
\frac{1}{(\epsilon_{n's} - \epsilon_{ns})
(\epsilon_{np} - \epsilon_{ns} - \omega)}
\right.
\nonumber \\ &&
\hspace*{3cm}
\left.
+ \frac{1}{(\epsilon_{n's} - \epsilon_{ns})
(\epsilon_{n's} - \epsilon_{np} + \omega)}
\right)
\,,
\end{eqnarray}
which vanishes exactly.
The contributions of the rest is
$\propto Q_w \langle n s | \vec{d}\cdot\vec{E} | n' p \rangle
\langle n'p | V_C| np \rangle$, as given below
in $F(\omega)$ of eq.(\ref{atomic and coulomb factor}).

\begin{figure*}[htbp]
 \begin{center}
 \epsfxsize=0.5\textwidth
 \centerline{\epsfbox{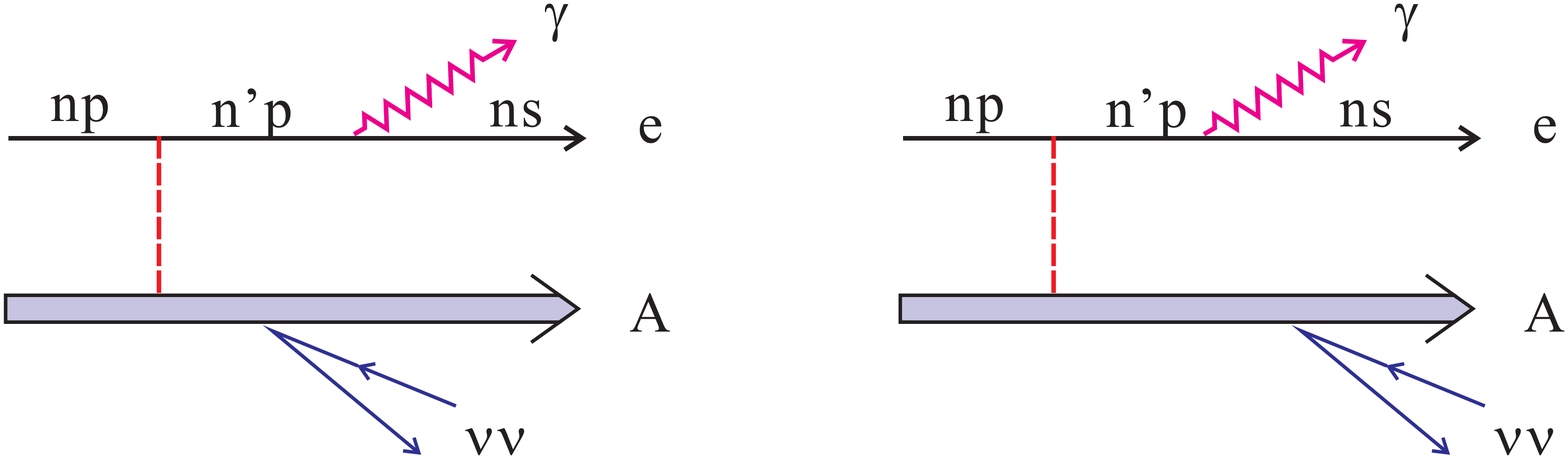}} \hspace*{\fill}
   \caption{RENP diagrams 2 for alkali atoms.
}
   \label{core renp pc 2}
 \end{center} 
\end{figure*}

\begin{figure*}[htbp]
 \begin{center}
 \epsfxsize=0.5\textwidth
 \centerline{\epsfbox{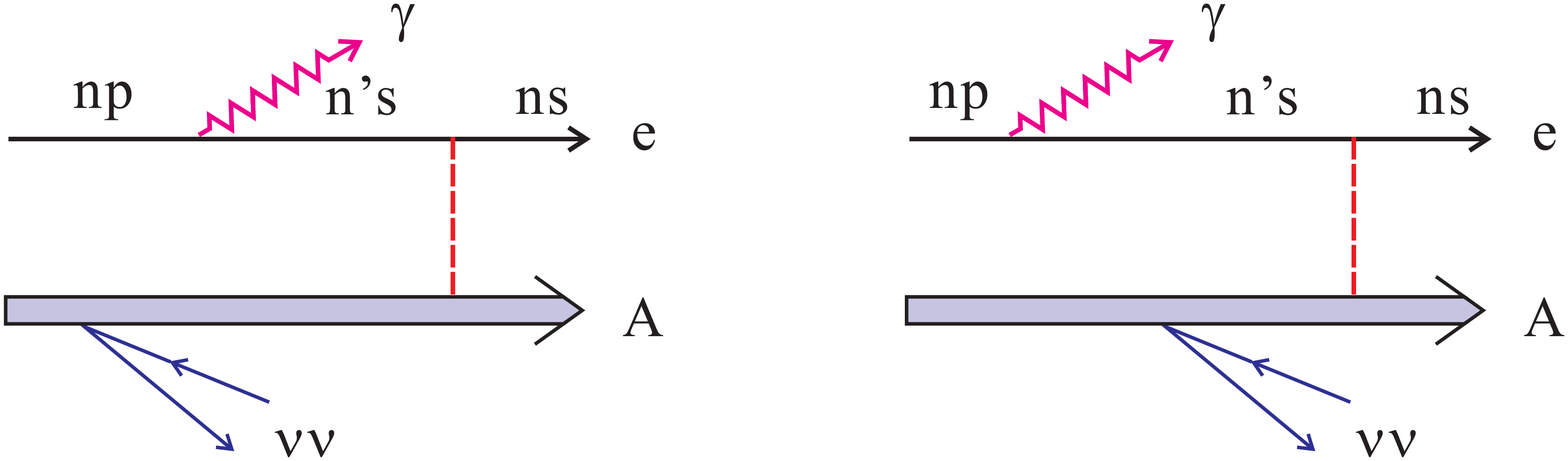}} \hspace*{\fill}
   \caption{RENP diagrams 3 for alkali atoms.
}
   \label{core renp pc 3}
 \end{center} 
\end{figure*}

RENP spectrum formula for alkali atomic transition
$|n' p \rangle \rightarrow |ns \rangle + \gamma +\nu \nu\,,
n' = n+1 $ is given by
\begin{eqnarray}
&&
\Gamma_{\gamma 2\nu}(\omega ;\, t) = \Gamma_{0}F^2(\omega) 
I(\omega)\eta_{\omega}(t)
\,, \hspace{0.5cm}
\Gamma_{0} = \frac{3}{4}G_F^2 n^3 V \epsilon_{eg} 
\,,
\\ &&
F(\omega) = 
\frac{ Q_w J_N (\epsilon(n'p) -  \epsilon(ns))}
{\epsilon(n'p) -  \epsilon(np)} 
\frac{1}{\sqrt{3\pi}}
\frac{d_{n'p ns}}{(\epsilon(n'p) -  \epsilon(np)+\omega)
(\epsilon(np) -  \epsilon(ns)- \omega) }
\,, \hspace{0.5cm}
d_{ab} = \sqrt{3\pi \frac{\gamma_{ab}}{(\epsilon_{a}- \epsilon_{b})^3}} 
\,,
\label{atomic and coulomb factor}
\\ &&
I(\omega) = 
\sum_{i}\Delta_{i}(\omega)
I_{i}(\omega) \theta(\omega_{ii}-\omega)
\,,\hspace{0.5cm}
\omega_{ii} =  \frac{\epsilon_{eg}}{2} - \frac{2m_i^2}{\epsilon_{eg}}
\,,
\label{rnpe spectrum rate}
\\ &&
I_{i}(\omega) =  
\frac{\omega^2}{3} + \frac{2m_i^2 \omega^2}{3 \epsilon_{eg}(\epsilon_{eg}-2\omega)} 
+ m_i^2 (1 + \delta_M)
\,,  \hspace{0.5cm}
\Delta_{i}(\omega) 
= \left( 1
 -  \frac{4 m_i^2}{\epsilon_{eg} (\epsilon_{eg} -2\omega) }
\right)^{1/2}
\,,
\label{rnpe spectrum rate 1}
\end{eqnarray}
with $\delta_M = 1$ for the Majorana case and zero for the Dirac case.
The relation between transition dipole $d_{ab}$ and transition rate
$\gamma_{ab}$ 
(A-coefficient) has been used.

The photon energy spectrum from RENP
is continuous below a threshold slightly below
the half of the energy difference of initial and final states,
$\epsilon_{eg}/2 - 2m_0^2/\epsilon_{eg} $ with $m_0$ the smallest neutrino mass, 
hence is separated from the familiar D1 line of alkali atoms at
$\epsilon_{eg}$.
The spontaneous (and not macro-coherent)
emission spectrum of two-photon decay  $|e\rangle \rightarrow |g\rangle
+ \gamma + \gamma$ is continuous starting from $\epsilon_{eg}/2$,
but has negligible rates.

The overall rate scale is given by $\Gamma_0$, which
has the dimension of mass, or $s^{-1}$, in our natural
unit of $\hbar = c = 1$.
Numerically, this value is 
\begin{eqnarray}
&&
\Gamma_0 
\sim 54 {\rm mHz} \frac{\epsilon_{eg}}{{\rm eV}} (\frac{n}{10^{21}{\rm cm}^{-3}})^3 
\frac{V}{10^2 {\rm cm}^3}
(\frac{100 {\rm MHz}}{{\rm eV}^3})^{-1}
\,.
\end{eqnarray}
As a reference parameter set, we took a target
number density $n= 10^{21}$cm$^{-3}$, a target volume $V= 10^2$cm$^3$, 
A-coefficients, $\gamma_{ab}$'s, in 100 MHz unit, and 
an available energy $\epsilon_{eg} = 1$eV, along with
all energies in the eV unit.
The rate dependence on these parameters is as explicitly indicated
in this equation.

The spectral shape given by this formula
is substantially different from the case of valence RENP in the preceding works
of the spin current \cite{ptep overview},
in particular in the low energy limit $\omega \rightarrow 0$.
The reason for this is in the nuclear mono-pole current
in the neutrino emission vertex, different from the spin current in the
valence RENP.
Calculation leading to the spectral rate $I(\omega)$
is sketched in Appendix.

The factor $\eta_{\omega}(t)$ is the extractable fraction of
field intensity $\epsilon_{eg}n$ stored in the initial upper level
$|e\rangle$. The storage and development of
target polarization is induced by two trigger laser irradiation
of $\omega + \omega' = \epsilon(n'p) - \epsilon(ns), \omega < \omega'$.
The storage is due to a second order QED process,
for instance M1$\times$E1 type of two-photon paired super-radiance (PSR) 
via virtual intermediate state $n'p_{1/2} \rightarrow n'p_{3/2}
\rightarrow ns_{1/2} $ in alkali atoms.
The calculation of $\eta_{\omega}(t)$ requires numerical solution of the master
equation for developing fields and target polarization
given in \cite{yst pra}, \cite{ptep overview}.
Usually, $\eta_{\omega}(t)$ is much less than unity,
and depends on experimental conditions.
In the present work we use a conservative 
value of $\eta_{\omega}(t)$ in the range $10^{-6}$
\cite{eta in ptep overview}.
The macro-coherent development of field at frequency $\omega$
and macroscopic polarization between $n'p$ and $ns$ up to several to
10 nano-second time range
is a prerequisite for experimental success of RENP.
The macro-coherence is expected to decay after
the phase relaxation time $T_2$.

\begin{figure*}[htbp]
 \begin{center}
 \epsfxsize=0.6\textwidth
 \centerline{\epsfbox{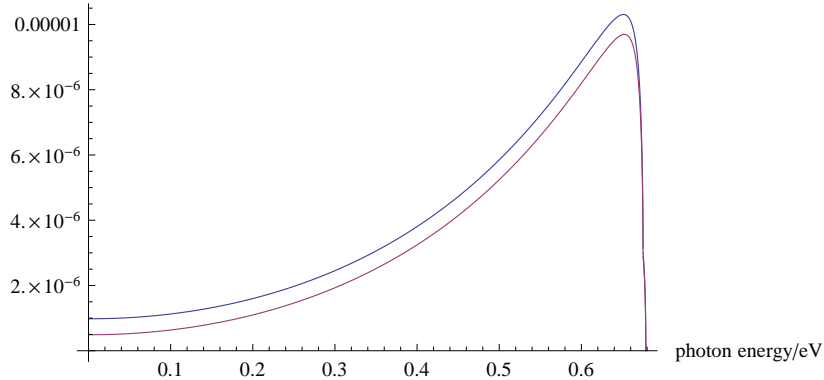}} \hspace*{\fill}
   \caption{Cs RENP spectrum from de-excitation of energy level at $ 6P_{1/2}(1.3859 {\rm eV})$,
assuming the smallest neutrino mass of 0.1 eV in the normal
hierarchical (NH) mass pattern, the Majorana case in blue and the 
Dirac case in magenda, taking other masses and mixing angles
consistent with neutrino oscillation experiments.
The actual Cs RENP rate is obtained by multiplying 
$1.5 \times 10^{5}(n/10^{21}{\rm cm}^{-3})^3( V/10^2{\rm cm}^3)
(\eta_{\omega}(t)/10^{-6})$Hz
}
   \label{cs overall spectrum}
 \end{center} 
\end{figure*}

\begin{figure*}[htbp]
 \begin{center}
 \epsfxsize=0.6\textwidth
 \centerline{\epsfbox{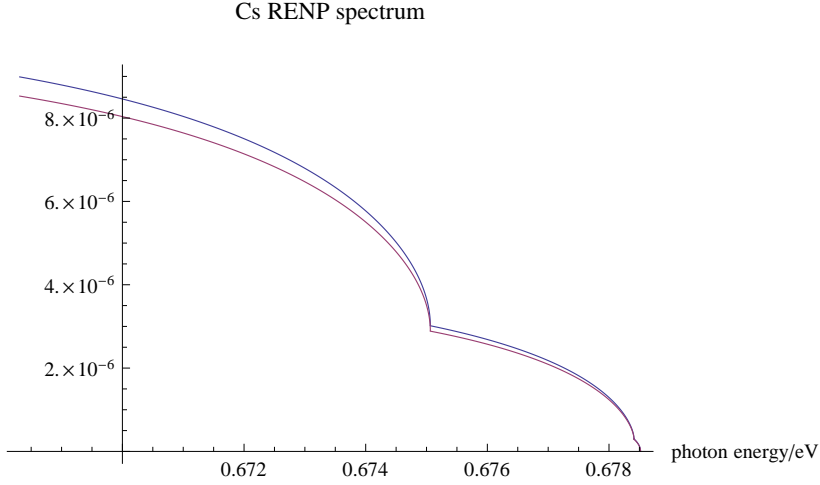}} \hspace*{\fill}
   \caption{Threshold region corresponding to Fig(\ref{cs overall spectrum})
}
   \label{cs threshold spectrum}
 \end{center} 
\end{figure*}

In Fig(\ref{cs overall spectrum}) and Fig(\ref{cs threshold spectrum})
 we plot the spectral shape for $^{133}$Cs.
Cs data used are
states $|e\rangle = 6P_{1/2}(1.3859 {\rm eV})\,, 
|g\rangle = 6S_{1/2} (0)$
and A-coefficient $7.93 \times 10^5$s$^{-1}$ for $7P_{1/2}(2.6986) \rightarrow 6S_{1/2}$,
taken from NIST \cite{nist}.
For the smallest neutrino mass as large as 0.1 eV as in this example
the Majorana vs Dirac distinction is possible by Cs RENP, but
for smaller mass values it becomes difficult requiring a large statistic
data of RENP.
The situation for MD distinction is improved for smaller
atomic spacings \cite{dpsty-plb}.

For another example we take Xe atomic de-excitation 
from 6s$^3$P$_1$ transition.
This is an electron-hole system consisting of a valence electron of 6s, 7s,6p and
a hole of 5p, much like two valence electron system.
We shall use a different scheme from that considered in \cite{ptep overview},
to utilize the nuclear mono-pole contribution.
Data used are energy levels of  6s$^3$P$_1$(8.437 eV) for initial state,
and  7s$^3$P$_1$(10.593 eV),
and its A-coefficient, 
$\gamma_{7s5p} = 8.51 \times 10^7$s$^{-1}$.
$^{131}$Xe RENP rate from nuclear pair emission is given by
\begin{eqnarray}
&&
\Gamma_{\gamma 2\nu}(\omega ;\, t) = \Gamma_{0}F_X^2(\omega) 
I(\omega)\eta_{\omega}(t)
\,, \hspace{0.5cm}
F_X (\omega) =
\frac{ Q_w J_N ( \epsilon_{7s} - \epsilon_{6s}) }{ \epsilon_{7s} - \epsilon_{5p}}
\frac{1}{\sqrt{3\pi}}
\frac{d_{7s 5p}}{(\epsilon_{7s} - \epsilon_{6s} + \omega) 
(\epsilon_{6s}- \epsilon_{5p} - \omega)} 
\,.
\label{xe renp rate}
\end{eqnarray}
Abbreviated notations are used, paying attention
to single electron transitions:
for instance $7s$ here means the atomic state $5p^5 7s$ and
$5p$ is an orbital in the closed shell $5p^6$ with $\epsilon_{5p}=0$
by definition of the energy origin.

\begin{figure*}[htbp]
 \begin{center}
 \epsfxsize=0.6\textwidth
 \centerline{\epsfbox{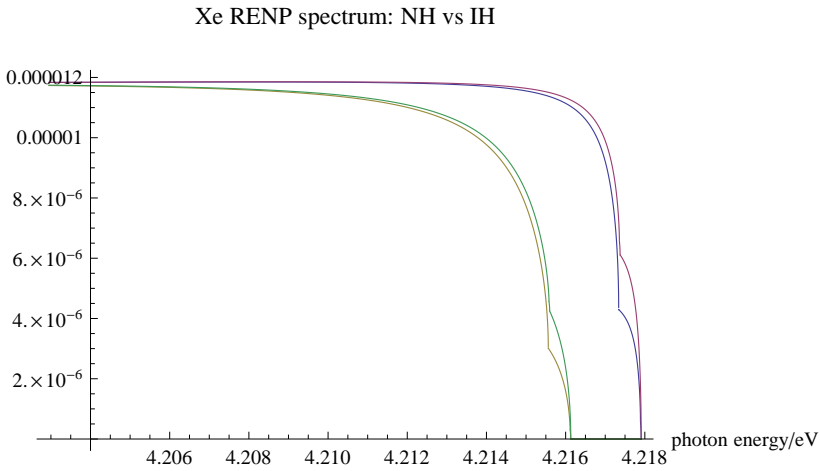}} \hspace*{\fill}
   \caption{Xe spectral shape for Dirac and Majorana RENP.
   Actual rate should be multiplied by $\sim 4 \times 10^3 $Hz for Xe gas
density of $7\times 10^{19} $cm$^{-3}$,
volume $10^2$cm$^3$, and $\eta_{\omega}= 10^{-6}$.
   Assumed smallest neutrino masses are 50, 100  
meV for the normal (NH) and inverted hierarchical (IH) patterns.
NH 50 meV  is depicted
in blue, IH 50 meV in magenda, NH 100 meV in brown, and
IH 100 meV in green. 
Majorana and Dirac cases are degenerate with this resolution.
     }
   \label{xe renp spectrum 1}
 \end{center} 
\end{figure*}

Its spectral shape is given in Fig(\ref{xe renp spectrum 1}), which
shows that neutrino mass differences of this size
and different hierarchical
mass patterns can be differentiated.
MD distinction is impossible with assumed neutrino masses.
Although the RENP rate is much smaller due to
the assumed atom density appropriate for gas target,
the gas target has a number of merits compared with
solid targets such as a larger phase relaxation time $T_2$.

\vspace{0.5cm}
{\lromn 6}
{\bf Summary}

\vspace{0.5cm}
We have presented a new enhancement mechanism of RENP due to
the mono-pole vertex of neutrino pair emission
from inner core electrons and nucleus in heavy atoms.
The enhancement factor for RENP rates is very large,
depending on the atomic number $\propto Z^{8/3}$
for pair emission from core electrons
and $\propto Q_w Z^{8/3}$ for pair emission from nucleus where
$Q_w \sim N - 0.044 Z$ is the electroweak neutral charge of nucleus.
Both rates and spectral shapes of emitted photon energy
have been calculated and examples of Cs and Xe RENP have been provided.
The new mechanism of mono-pole current
opens a variety of possibilities in selection of ideal RENP targets.

\vspace{0.5cm}
{\lromn 7}
{\bf Appendix: Coulomb integral in Thomas-Fermi model}

\vspace{0.5cm}
In the Thomas-Fermi model \cite{atomic physics}
one assumes the degenerate Fermi gas of electrons
at each local point of atom, and relates the Fermi momentum to
the number density.
The kinetic energy at the Fermi momentum is balanced to the potential
energy exerted to electron.
In another word, the pressure gradient of degenerate gas
is balanced against the electrostatic potential.
This gives a relation of the electron number density $n_e(r)$ to
the potential $\varphi(r)$:
\begin{eqnarray}
&&
\frac{(3\pi^2 n_e(r))^{2/3}}{2m_e} - e \varphi(r) = 0
\,.
\end{eqnarray}
The spherical symmetry is assumed.

The second important equation is the Poisson equation,
relating the electron number density to the potential.
Combined with the density-potential relation above,
one arrives at a self-consistent equation for the potential
\begin{eqnarray}
&&
 \frac{1}{r} \frac{d^2}{dr^2} (r\varphi) = \frac{e}{3\pi^2} (2m_e e \varphi)^{3/2}
 \,.
\end{eqnarray}
It is convenient to introduce dimensionless units of
\begin{eqnarray}
&&
\chi = \frac{4\pi}{Ze}r \varphi
\,, \hspace{0.5cm}
r = b x
\\ &&
b = \frac{(3\pi)^{2/3}}{2^{7/3}}  \frac{Z^{-1/3}}{\alpha m_e} 
\sim 0.8853  Z^{-1/3}a_B
\,.
\end{eqnarray}
The Thomas-Fermi equation is written for $\chi(x)$;
\begin{eqnarray}
&&
x^{1/2} \frac{d^2 \chi}{d^2 x} = \chi^{3/2}
\,.
\end{eqnarray}
The asymptotic behavior with $x \rightarrow \infty$ is worked out, to give
$ \chi(x) \rightarrow 144/x^3$.
The boundary condition at the origin is set from the physical setup,
the nuclear charge, which dictates
$\chi(x) \rightarrow 1$ as $x\rightarrow 0$, along with $\chi'(0) = 0$.
The problem thus becomes an eigen-value problem.
The eigen-function satisfies $\chi(x) = 1 - c_1 x +\cdots,
c_1 \sim 1.588 $ as $x \rightarrow 0$ \cite{atomic physics}.
The electron number density is given by
\begin{eqnarray}
&&
n_e = \frac{32}{9\pi^3} Z^2 (\alpha m_e)^3 (\frac{\chi}{x})^{3/2}
\,.
\end{eqnarray}

The Coulomb interaction between a valence electron and
all core electrons in the closed shell is given by
\begin{eqnarray}
&&
\int d^3 r_1 d^3 r_2 |\psi_n(\vec{r}_1)|^2 n_e(\vec{r}_2)
\frac{\alpha}{|\vec{r}_1 - \vec{r}_2|} \equiv  J_C \,.
\end{eqnarray}
We may assume that dependence of this quantity on the quantum number of
valence electron $n$ is weak and define the Coulomb integral as $\alpha J$.
This quantity is given in dimensionless units,
\begin{eqnarray}
&&
 J_C = \frac{(4\pi)^2}{(6\pi^4)^{2/3}} Z^{4/3} \frac{1}{2} \alpha^2 m_e {\cal J}
\sim 31 {\rm eV}  Z^{4/3}  {\cal J}
\,,
\\ &&
{\cal J} = \int_0^{\infty} dx_1 x_1^{-1/2} \left( \chi(x_1) \right)^{3/2} 
\int_0^{x_1} dx_2 x_2^{1/2} \left( \chi(x_2) \right)^{3/2} 
\,.
\end{eqnarray}
Value of ${\cal J} \sim 0.23$ is obtained
by numerically solving Thomas-Fermi equation
and by integrating results, to give
\begin{eqnarray}
&&
(J_C)^2 \sim 50  Z^{8/3} {\rm eV}^2
\,.
\end{eqnarray}

Another important integral used in the text is
the Coulomb integral between valence electron and
nucleus, which is
\begin{eqnarray}
&&
Z\alpha \int d^3 r_1 d^3 r_2 
\frac{|\psi_N(\vec{r}_2)|^2 n_e(\vec{r}_1)}{|\vec{r}_1 - \vec{r}_2|}
\sim \alpha \int d^3 r \frac{n_e(r)}{r}
\,,
\end{eqnarray}
in the small nucleus limit.
Estimate of this quantity in the Thomas-Fermi model is
\begin{eqnarray}
&&
\frac{2^{7/3}}{(3\pi)^{2/3}} Z^{4/3} \alpha^2 m_e \int_0^{\infty} dx
\frac{\chi(x)^{3/2}}{x^{1/2}} \sim 31 {\rm eV} Z^{4/3}
\,.
\end{eqnarray}

\vspace{0.5cm}
{\lromn 8}
{\bf Appendix: Phase space integral over neutrino momenta}

\vspace{0.5cm}
We start from two neutrino emission vertex (\ref{2nu vertex 2}),
its square to be multiplied by E1 photon emission factor
$(\vec{e}\cdot\vec{E})^2$ from valence electron
and by the Coulomb factor $F$ of eq.(\ref{coulomb factor}) for rates.
Here we concentrate on summation over helicities and momenta of
two emitted neutrinos.

Using the helicity summation formula of \cite{my-prd}, 
\begin{eqnarray}
&&
\sum_{h_i} |j_{\nu}\cdot j^{e,q } A|^2 = 
\frac{1}{2} (1 + \frac{\vec{p}_1\cdot\vec{p}_2}{E_1E_2}+\delta_M 
\frac{m_1 m_2}{E_1E_2} ) 
j_0^{e,q} (j_0^{e,q})^{\dagger}|A|^2
+ \cdots
\,,
\label{mono-pole currents}
\end{eqnarray}
where $j_0^{e,q}$ is the zero-th component electron current,
either of electron or of quark,
and $(E_i, \vec{p}_i)$ are neutrino 4-momenta.
The function $A$ refers to all the rest of amplitudes including
QED vertex, energy denominators, and all coupling constants.
In previous works on valence RENP, the 3-vector part $\propto \vec{j}_e$
of electron current (spin-current), 
\begin{eqnarray}
&&
\sum_{h_i} |j_{\nu}\cdot j^e A|^2 = 
\frac{1}{2} (1 - \frac{\vec{p}_1\cdot\vec{p}_2}{E_1E_2} -\delta_M 
\frac{m_1 m_2}{E_1E_2} ) 
\vec{j}_e\cdot\vec{j}_e^{\dagger} |A|^2 +  \frac{\vec{p}_1\cdot\vec{j}^e \vec{p}_2\cdot\vec{j}^e  |A|^2}{E_1 E_2}
+ \cdots
\,,
\label{mono-pole currents 2}
\end{eqnarray}
has been relevant.
Difference of the sign $\pm \frac{\vec{p}_1\cdot\vec{p}_2}{E_1E_2}$
appears in the suppressed region of the spectrum:
for the mono-pole current (\ref{mono-pole currents})
the low energy limit $\omega \sim 0$
neutrino momenta are nearly balanced, $\vec{p}_1 \sim - \vec{p}_2$,
and there is a more suppression in the low energy limit
for the mono-pole case.
In eqs.(\ref{mono-pole currents}) and .(\ref{mono-pole currents 2})
we neglected possibly time reversal odd terms.

In the phase space integral of neutrino momenta,
\begin{eqnarray}
&&
\int \frac{d^3 p_1 d^3 p_2}{(2\pi)^2} \delta(E_1 + E_2 + \omega - \epsilon_{eg}) 
\delta(\vec{p}_1 + \vec{p}_2 + \vec{k}) (\cdots)
\end{eqnarray}
one of the momentum integration is used to eliminate the
delta function of the momentum conservation.
The resulting energy-conservation is used to fix the relative angle
factor $\cos \theta $
between the photon and the remaining neutrino momenta, 
$\vec{p}_1 \cdot \vec{k} = p_1 \omega \cos \theta$.
Noting the Jacobian factor $E_2/p\omega$
from the variable change to the cosine angle, one obtains one dimensional integral
over the neutrino energy $E_1$:
\begin{eqnarray}
&&
I_{ij}(\omega) \frac{\Delta_{ij}(\omega)}{2\pi }
\equiv 
\frac{1 }{2\pi \omega}
\int_{E_-}^{E_+} d E_1E_1 E_2 \frac{1}{2} (1 + \frac{\vec{p}_1\cdot\vec{p}_2}{E_1E_2}+\delta_M 
\frac{m_1 m_2}{E_1E_2} ) 
\,, \hspace{0.5cm}
E_2 = \epsilon_{eg} - \omega - E_1
\,.
\end{eqnarray}
The angle factor constraint $|\cos \theta| \leq 1$
places a constraint on the range of neutrino energy integration,
\begin{eqnarray}
&&
E_{\pm} =
\frac{1}{2} \left( (\epsilon_{eg} - \omega) (1 +
\frac{m_i^2 - m_j^2}{\epsilon_{eg}(\epsilon_{eg} - 2\omega)} )
\pm \omega \Delta_{ij}(\omega)
\right)
\,,
\\ &&
\Delta_{ij}(\omega) 
= \left\{
\left(1 - \frac{ (m_i + m_j)^2}{\epsilon_{eg} (\epsilon_{eg} -2\omega) } \right)
\left(1 - \frac{ (m_i - m_j)^2}{\epsilon_{eg} (\epsilon_{eg} -2\omega) } \right)
\right\}^{1/2}
\,.
\end{eqnarray}
The integrand is a quadratic function of neutrino energy
\cite{dpsty-plb},
and it is easily integrated to give
\begin{eqnarray}
&&
\hspace*{-1cm}
I_{ij}(\omega) =  
\frac{\omega^2}{3} + \frac{1}{2}(m_i^2 + m_j^2)
+ \frac{1}{3} \frac{\omega^2 (m_i^2 + m_j^2)}{\epsilon_{eg} (\epsilon_{eg} -2\omega) }
- 
\frac{3}{4} \frac{(\epsilon_{eg} - \omega)^2}
{\epsilon_{eg}^2(\epsilon_{eg} - 2\omega)^2} (m_i^2 - m_j^2)^2
 +\delta_M m_i m_j 
\,.  
\end{eqnarray}
The result, eq.(\ref{rnpe spectrum rate 1}) and
in other places of the text, $I_i(\omega) = I_{ii}(\omega)$
is needed.

\vspace{0.3cm}
{\bf Acknowledgements}
\hspace{0.2cm}
We appreciate M. Tanaka for a discussion.
This research was partially supported by Grant-in-Aid for Scientific
Research on Innovative Areas "Extreme quantum world opened up by atoms"
(21104002)
from the Ministry of Education, Culture, Sports, Science, and Technology.


\begin{thebibliography}{99}
\bibitem{eta in ptep overview}
In \cite{ptep overview} a result for
numerical simulation of $\eta_{\omega}(t)$
is presented for pH$_2$ molecule target
(strong source of paired super-radiance (PSR)
of E1 $\times$ E1 transition, and  see
Fig 14 of this reference for time dependence).
Its time dependence is complicated:
a fast rise in $O(2$ ns), then a plateau region
of magnitude $O(10^{-2} \sim 10^{-3})$ of duration of
several nano-seconds, finally gradual decrease
ending around $10^{-6}$ at $\sim$ 12 ns (end time of calculation).

For RENP rate calculations,
numerical simulations based on the master
equation given in \cite{ptep overview} should be
performed for weaker PSR process of 
specific targets considered,
which is expected to give different time
profile and larger values of $\eta_{\omega}(t)$.
\end{thebibliography}
\end{document}